\documentclass[aps,prl,twocolumn]{revtex4-2}

\usepackage{graphicx}
\usepackage{dcolumn}
\usepackage{bm}
\usepackage{amsmath}
\usepackage{amssymb}
\usepackage{url}
\usepackage{color}
\usepackage[colorlinks,bookmarks=false,citecolor=darkblue,linkcolor=red,urlcolor=blue]{hyperref}

\definecolor{darkred}{rgb}{0.7,0.0,0.0}
\definecolor{darkblue}{rgb}{0,0.02,0.45}
\usepackage{float}

\graphicspath{{figures/}}

\begin{document}

\title{Reinforcement of superconductivity by quantum critical fluctuations of metamagnetism in UTe$_2$}


\author{Y. Tokiwa}
\email[]{*tokiwa.yoshifumi@jaea.go.jp}
\affiliation{%
ASRC,
Japan Atomic Energy Agency
Tokai, Ibaraki 319-1195, Japan
}%

\author{P. Opletal}
\affiliation{%
ASRC,
Japan Atomic Energy Agency
Tokai, Ibaraki 319-1195, Japan
}%

\author{H. Sakai}
\affiliation{%
ASRC,
Japan Atomic Energy Agency
Tokai, Ibaraki 319-1195, Japan
}%

\author{S. Kambe}
\affiliation{%
ASRC,
Japan Atomic Energy Agency
Tokai, Ibaraki 319-1195, Japan
}%

\author{E. Yamamoto}
\affiliation{%
ASRC,
Japan Atomic Energy Agency
Tokai, Ibaraki 319-1195, Japan
}%

\author{M. Kimata}
\affiliation{%
IMR, Tohoku University, Sendai, Miyagi, 980-8577, Japan
}%

\author{S. Awaji}
\affiliation{%
IMR, Tohoku University, Sendai, Miyagi, 980-8577, Japan
}%

\author{T. Sasaki}
\affiliation{%
IMR, Tohoku University, Sendai, Miyagi, 980-8577, Japan
}%

\author{D. Aoki}
\affiliation{%
IMR, Tohoku University, Ibaraki 311-1313, Japan
}%

\author{Y. Haga}
\affiliation{%
ASRC,
Japan Atomic Energy Agency
Tokai, Ibaraki 319-1195, Japan
}%

\author{Y. Tokunaga}
\affiliation{%
ASRC,
Japan Atomic Energy Agency
Tokai, Ibaraki 319-1195, Japan
}%


\date{\today}

\begin{abstract}
The normal-conducting state of the superconductor UTe$_2$ is studied by entropy analysis for magnetic fields along the $b$-axis, obtained from magnetization using the relation $(\partial M/\partial T)_B=(\partial S/\partial B)_T$. We observe a strong increase in entropy with magnetic field due to metamagnetic fluctuations (spatially uniform, $Q=0$). The field dependence is well described by the Hertz-Millis-Moriya theory for quantum criticality of itinerant metamagnetism. Notably, the lower bound of the quantum-critical region coincides well with the position of the minimum in the superconducting transition temperature $T_c(B)$. Hence, our results suggest that $Q=0$ fluctuations reinforce the superconductivity.

\end{abstract}

\maketitle
Unconventional superconductivity arises from an anomalous normal-conducting (NC) state~\cite{mathur-nature-98,Cooper2009,Hashimoto2012}. The NC states of heavy-fermion, iron-pnictide, and cuprate high-$T_{\rm c}$ superconductors exhibit non-Fermi liquid behavior with unusual $T$-dependence of resistivity due to quantum fluctuations from magnetic quantum criticality. Extensive studies have established close relationships between superconductivity and magnetic criticality, leading to the widely accepted belief that the superconducting (SC) pairing interaction is provided by such fluctuations. Hence, investigating the NC states of unconventional superconductors is of great importance to gain insights into the SC pairing mechanism.

In this study, our focus is on the recently discovered superconductor, UTe$_2$, which undergoes a superconducting transition at $T_{\rm c}=1.6$ K~\cite{Ran2019,Aoki2019}. The superconductivity has garnered significant attention due to its reinforcement (re-entrance) under high magnetic fields and the presence of multiple superconducting phases~\cite{Ran2019,Knebel2019,Aoki2020a,Kinjo2022,Rosuel2022,Sakai2023}. The pairing symmetry has been subject to extensive investigations~\cite{Jiao2020,Kittaka2020,Xu2019,Ishizuka2019,Machida2021,Kanasugi2022}. Furthermore, considerable effort has been dedicated to improving the crystal quality, leading to a significant enhancement of $T_c$ to over 2 K from the originally reported value~\cite{Haga2022,Rosa2022a,Aoki2022a,Sakai2022}. Notably, recent progress has allowed the successful observation of quantum oscillations in high-quality single crystals grown using the molten-salt flux (MSF) method~\cite{Aoki2023,Eaton2023}.

The reinforcement of superconductivity is observed when the field is applied along the hard $b$-axis, where a sharp metamagnetic (MM) transition takes place around 35\,T~\cite{Knafo2019,Miyake2019}. There, the superconducting transition temperature $T_{\rm c}$ is initially suppressed with magnetic field; however, it remarkably enhances above $B^{\star}\sim$15,T. As a result, a minimum in $T_{\rm c}$($B$) is observed as a function of magnetic field. For certain magnetic field angles between the $b$- and $c$-axes, the magnetic field completely suppresses the superconductivity, but above 40,T, the superconductivity reappears~\cite{Ran2019a,Knebel2019}. These reinforcement and reappearance bear resemblance to other uranium ferromagnetic superconductors, URhGe and UCoGe of which the superconductivity may be mediated by ferromagnetic fluctuations~\cite{Aoki2019a,Aoki2001,Huy2007,Hattori2012,Tokunaga2015}.
Unlike these superconductors, UTe$_2$ is paramagnetic. The absence of ferromagnetism and the similarity to ferromagnetic superconductors open the intriguing possibility of ferromagnetic quantum fluctuations providing the superconducting pairing interaction. However, the magnetic susceptibility of recently grown high-quality crystals tends to saturate at low temperatures~\cite{Sakai2022}, which is consistent with a temperature-independent Knight shift~\cite{Tokunaga2022}. Instead, magnetic excitations at incommensurate $Q$-vectors have been observed by neutron scattering experiments~\cite{Duan2021,Knafo2021a}. Therefore, the nature of magnetic fluctuations responsible for the superconductivity remains unknown~\cite{Butch2022}. A broad anomaly observed in the NC state around 15 K in several thermodynamic, transport, and NMR experiments may be related to the key magnetic fluctuations~\cite{Tokunaga2019a,Eo2021,Willa2021,Tokunaga2022}.

In strong magnetic fields along the $b$-axis, fluctuations associated with the MM transition likely play a crucial role. Metamagnetism is identical to ferromagnetism in the sense that they both correspond to a spatially-uniform ($Q=0$) magnetic instability, as evidenced by the discontinuity in uniform magnetization, $M(Q=0)$, at the transition field. The only distinction from ferromagnetism lies in its induction by magnetic field. When the critical end point (CEP) of metamagnetism is tuned to zero temperature, the quantum fluctuations lead to the formation of an anomalous metallic state~\cite{millis02,Zacharias13}. Even when a system is not precisely tuned to QCEP but is in close proximity with the CEP at a finite temperature, the quantum-critical fluctuations can still induce anomalous state.~\cite{millis02,Gegenwart06,Zacharias13,Rost-Science09,tokiwa13,Aoki2011,Kotegawa2011}. In UTe$_2$, an increase in electron mass and longitudinal fluctuations with the field towards the metamagnetism is observed and is proposed to drive the reinforcement of superconductivity~\cite{Knafo2019,Knafo2021,Rosuel2022,Tokunaga2023}.

Entropy is a direct thermodynamic measure of fluctuations, as it accumulates at a quantum critical point and exhibits a peak as a function of control parameters, such as magnetic field and pressure~\cite{Garst2005,Wu2011}. Because unconventional superconducting pairing is believed to be mediated by such quantum critical fluctuations, it is interesting to map entropy in some parameter space. In UTe$_2$, the superconductivity couples to the magnetic field in a very peculiar manner, as evidenced by the non-monotonic behavior of $T_c(B)$. Therefore, the magnetic field-temperature parameter space is the most interesting to map entropy. To achieve this, we performed high-resolution magnetization measurements for $B\parallel b$-axis, enabling us to extract the field dependence of entropy in the NC state of UTe$_2$ using a thermodynamic relation, $(\partial M/\partial T)_B=(\partial S/\partial B)_T$. The consistency with specific heat \cite{Rosuel2022} is checked by comparing $C/T$ and $\int(dM/dT)/TdB$=$\int(dS/dB)/TdB$=$\gamma$ at low temperatures in Fermi liquid regime, where $\gamma$ is the Sommerfeld coefficient. (see Supplementary Material \cite{SM}). In this study, we reveal an anomalous increase in fluctuations attributed to metamagnetism in an entropy map. We investigate the nature of these fluctuations by fitting their field dependence with the Hertz-Millis-Moriya theory, which pertains to the quantum criticality of itinerant metamagnetism~\cite{millis02,Zacharias13}. The theory, widely recognized as a standard, is frequently employed to describe the anomalous behavior associated with metamagnetic quantum criticality~\cite{Gegenwart06,tokiwa13,Aoki2011,Weickert2010,Michor04}.

Magnetization is measured by a vibration-sample magnetometer (VSM) with an accuracy of $\sim$2$\times$10$^{-4}$\,emu in magnetic fields up to 24\,T and for temperatures between 2.2\,K and 65K, in the High-Field Laboratory for Superconducting Materials at the Institute for Materials Research at Tohoku University.
We measured magnetization $M$($B$) at various temperatures with a typical interval of 1\,K up to 20\,K, 2K up to 40\,K and 5\,K up to 65\,K to obtain a magnetization landscape~(see Supplementary Material~\cite{SM}). From the slope along $T$-axis we have a landscape of field derivative of entropy $\partial M$/$\partial T$=$\partial S$/$\partial B$. Then, $\partial S$/$\partial B$ is integrated over $B$ to obtain entropy increment $\Delta S$ from zero field, namely $\Delta S=\int_0^B(\partial M/\partial T)dB$.  Note that $\Delta S$ does not contain entropy at zero field. A large single crystal with a weight of 94\,mg, grown by the chemical vapor transport (CVT) method, is used. The SC transition temperature, the residual resistivity ratio and the residual specific heat coefficient divided by the value at the normal-conducting state of the same batch are respectively $T_c$=1.7 K, RRR=25 and $\gamma(0)/\gamma_{\rm N} = 0.43$. We note that the properties of the normal-conducting state remain unaffected by the sample quality. Specifically, the characteristics of magnetic fluctuations and the metamagnetic critical field show no variation across crystals of different quality~\cite{Tokunaga2023,Wu2023_arxiv}. We will delve into the discussion later, explaining how this robustness in the properties of metamagnetism leads to the independence of $B^\star$, where $T_c(B)$ exhibits a minimum, from variations in the sample quality.

\begin{figure}
\centering
\includegraphics[width=0.9\linewidth]{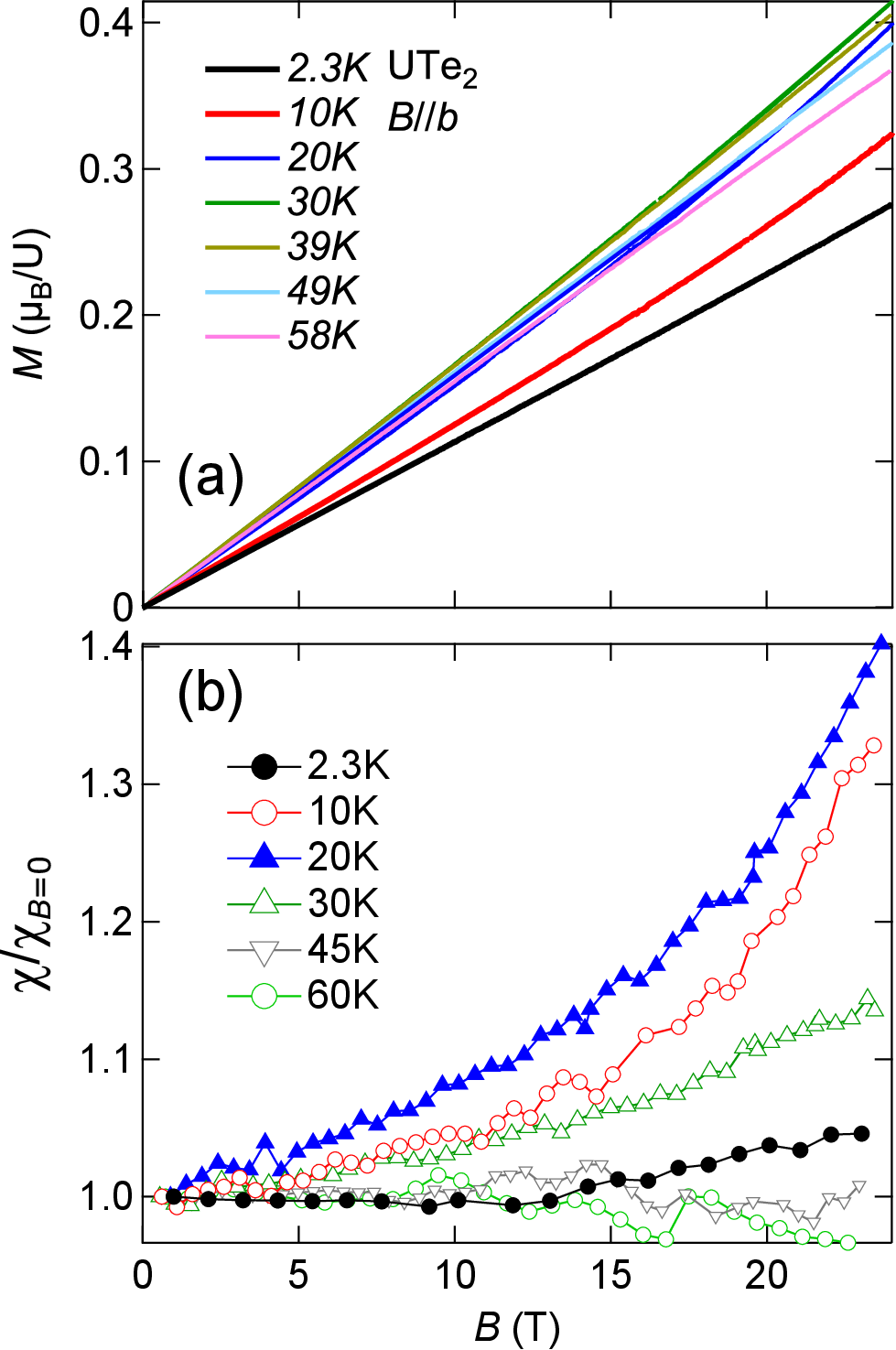}
\caption{(a) Magnetization of UTe$_2$ for the field along $b$-axis at different temperatures. (b) Magnetic field dependence of differential magnetic susceptibility normalized by the zero field values, $\chi/\chi_{B=0}$, of UTe$_2$ at different temperatures. }
\label{chi}
\end{figure}

Figure~\ref{chi}(a) shows the magnetization of UTe$_2$ for the field along $b$-axis. Magnetization at 10 and 20 K shows clearly the stronger-than linear field dependence due to metamagnetic fluctuations, which we will discuss later. Note that the magnetization of a paramagnet, such as those described by the Brillouin function, exhibits a downward curvature. As a consequence, the differential magnetic susceptibility ($\chi=dM/dB$) normalized by the zero-field value of a paramagnet is always less than 1 under magnetic fields, $\chi/\chi_{B=0}<1$. Figure ~\ref{chi}(b) displays such a normalized magnetic susceptibility of UTe$_2$ for the field along $b$-axis. Reflecting the non-linear magnetization, $\chi$ at 10 and 20 K increases with magnetic field. The non-linear magnetization ($\chi>$1) is observed up to 30 K, indicating the influence of metamagnetic fluctuations in a wide temperature range. Notably, $\chi$ at 10 and 20 K changes its slope above $\sim$15 T. At 2.3\,K, $\chi$ is nearly constant below $\sim$15 T, but, it deviates from the constant value above $\sim$15 T.

\begin{figure}
\centering
\includegraphics[width=\linewidth]{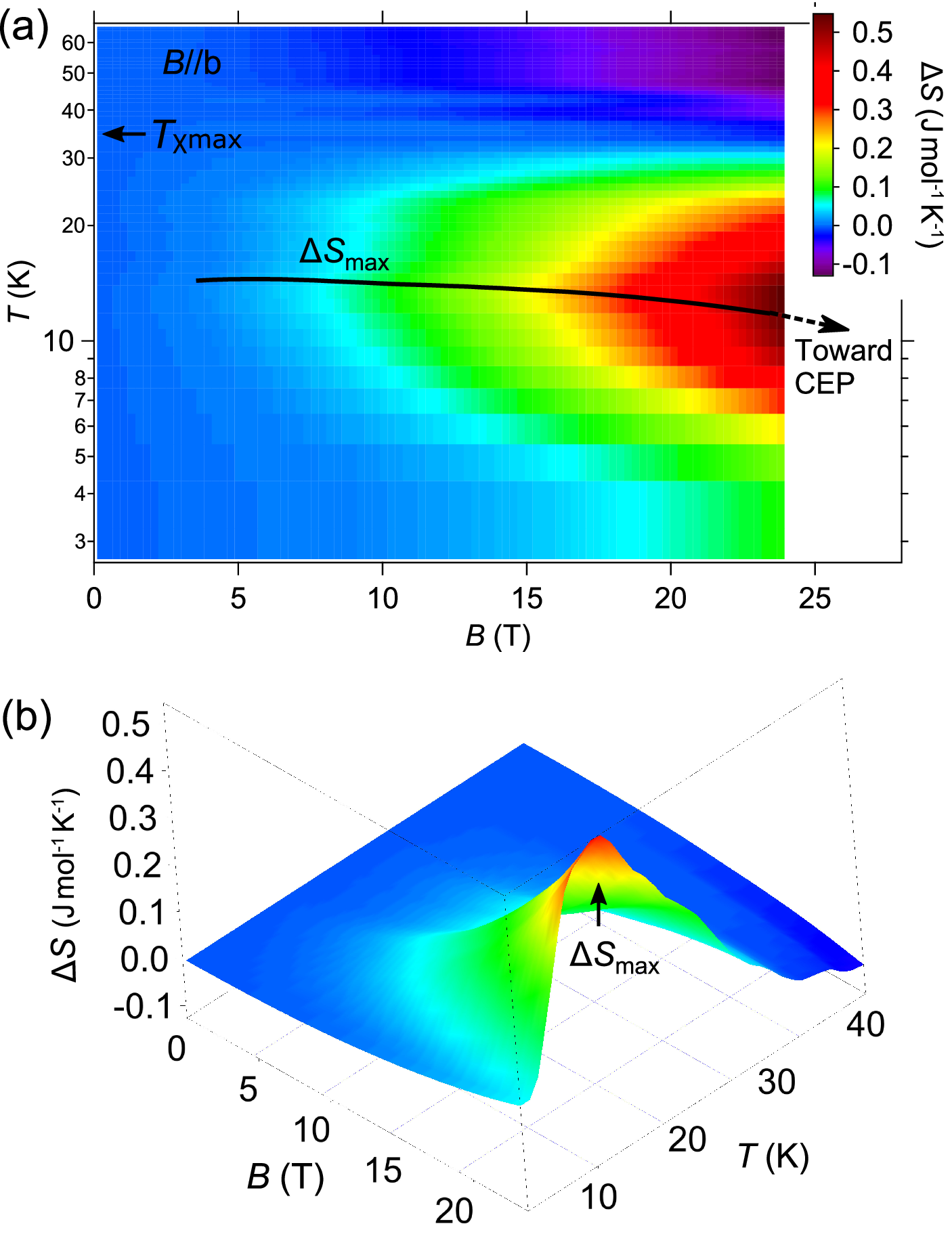}
\caption{\label{contour} (a) Color contour plot for magnetic entropy increment $\Delta S$ of UTe$_2$ in $T$-$B$ phase diagram for $B\parallel$b-axis. Solid black line corresponds to a maximum of $\Delta S(T)$ as a function of temperature. "CEP" denotes a critical end point for metamagnetic transition. $T_{\chi max}$ denotes a temperature, where magnetic susceptibility at zero field shows a maximum~\cite{Ikeda2006,Ran2019,Aoki2019,Miyake2019,Li2021}. (b) Three-dimensional plot of $\Delta S$ as a function of magnetic field and temperature. The arrow denotes the position of $\Delta S$ maximum in the temperature dependence at the highest magnetic field of 24 T.}
\end{figure}

From the $M$($B$) measurements, we obtain magnetic entropy increment $\Delta S$, as shown in Fig.\,\ref{contour}. For an ideal paramagnet entropy decreases with magnetic field, because magnetic field aligns magnetic moments. In heavy fermion compounds, there is a cross-over across $T_{\rm K}$~\cite{Tokiwa2023arxiv}. At high temperatures above $T_{\rm K}$ paramagnetic behavior is expected, while below $T_{\rm K}$ entropy depends on subtle field variations of density of states, since in Fermi liquid (FL) $S$=$\gamma T$.

In UTe$_2$, the paramagnetic behavior with decreasing entropy is found at temperatures above $\sim$35\,K, as shown in Fig.\,\ref{contour}. This temperature 35\,K corresponds to a maximum of magnetization as a function of temperature, where $\partial M/\partial T$=$\partial S/\partial B$=0, in agreement with the reported maximum of magnetic susceptibility around $T_{\chi max}\sim$35\,K~\cite{Ran2019,Miyake2019}. Across this temperature $\partial S/\partial B$ changes its sign.  Below this temperature $M$ decreases with decreasing $T$, therefore entropy increases with magnetic field, $\partial M/\partial T$=$\partial S/\partial B>$0. Such entropy increase is commonly observed in compounds with MM transitions/cross-overs due to increasing MM fluctuations toward the critical field~\cite{Rost-Science09,tokiwa13,Tokiwa2016}. In Figs.\,\ref{contour}(a,b) we plot a position of $\Delta S_{\rm max}$, where $\Delta S$ is the most as a function of temperature.(See also Supplementary Material for the temperature dependence of $\Delta S$ at different magnetic fields~\cite{SM}.) It shifts to lower temperatures with magnetic field, approaching the CEP of metamagnetism. At lower tempratures below the CEP, the entropy increase with magnetic field weakens. This is attributed to the weakening of magnetic fluctuations caused by the sharpening of metamagnetic transition with decreasing temperature. We note that the entropy increase still remains finite even at low temperatures, as can be seen in Fig.3(a), likely playing a crucial role in the reinforcement of superconductivity. Importantly, as demonstrated by Figure.\,\ref{contour}, the entropy increase at low temperatures is caused by the metamagnetism, which represents a $Q = 0$ instability.

\begin{figure}
\centering
\includegraphics[width=\linewidth]{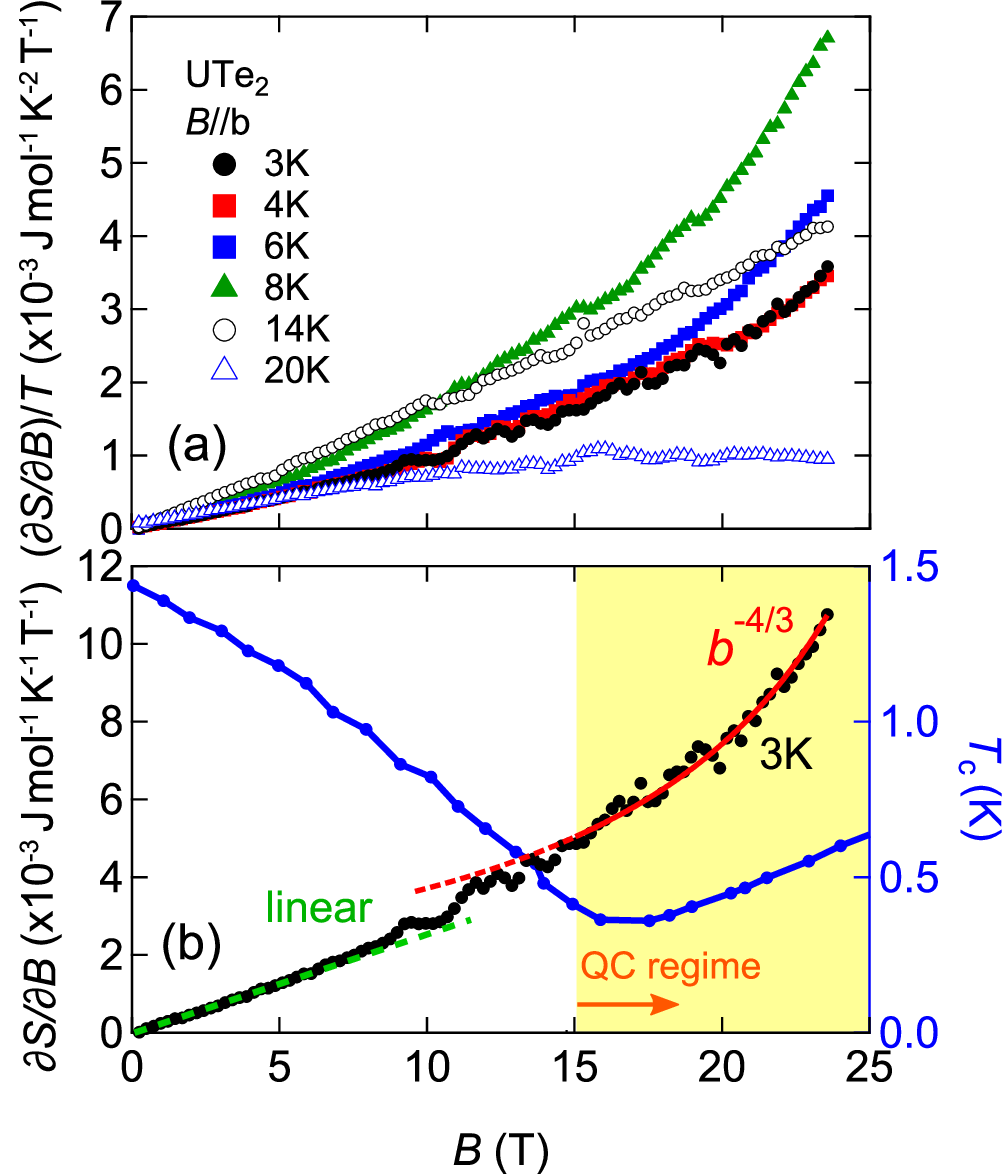}
\caption{\label{b-axis} (a) Magnetic-field derivative of entropy divided by temperature, $(\partial S/\partial B)/T$, of UTe$_2$ as a function of magnetic field along $b$-axis at different temperatures. (b) Comparison between magnetic-field derivative of entropy (left axis, black circles) and the superconducting transition temperature $T_c$ (right axis, blue circles)~\cite{Knebel2019}. Green dotted linear line is a guide for eye. Red solid line represents $b^{-4/3}$ field dependence, where $b$=$|B-B_c|$ and $B_c$=34.75\,T is the metamagnetic critical field~\cite{Rosuel2022}. "QC regime" denotes the field range of dominating quantum critical component where entropy follows the theoretical power-law for metamagnetic quantum criticality of two-dimension (red solid line)~\cite{Zacharias13}.}
\end{figure}

In Fig.~\ref{b-axis}(a), we plot $(\partial M/\partial T)/T=(\partial S/\partial B)/T$. In FL state, where $S=C=\gamma T$, this quantity equals to the field derivative of Sommerfeld coefficient, $\partial \gamma/\partial B$, which is $T$-independent. The data sets at 3 and 4 K lay on top of each other, indicating formation of  FL state below 4 K. They exhibit the upward curvature, indicating an accelerating increase of entropy at high fields. The upward curvature persists up to $\sim$8 K and changes to a downward curvature at high temperatures above the line of $\Delta S_{\rm max}$. Magnetic field derivative of entropy, $\partial S/\partial B$, at 3\,K in the FL regime, is plotted in Fig.\,\ref{b-axis}(b). At low fields it is linear in field, as expected from a symmetry consideration, $\partial S/\partial B$($B$)$=-\partial S/\partial B$(-$B$) with a leading $B$-linear term. The data gradually deviates from the linear dependence and increases rapidly with a positive curvature at high fields.

According to the Hertz-Millis-Moriya theory~\cite{millis02,Zacharias13}, $\partial S$/$\partial B$ in the critical regime follows $b^{(d-6)/3}$, where $b$=$|B$-$B_c|$ is the distance to the critical field, $B_c$, and $d$ is the dimensionality. Since Fermi surfaces consist of mainly cylindrical ones~\cite{Aoki2022c,Aoki2023,Eaton2023}, $d$=2 is appropriate, resulting in a power-law dependence $b^{-4/3}$. By fixing $B_c$=34.75\,T of a recently reported detailed study~\cite{Rosuel2022}, the data is well described by $b^{-4/3}$ above 15\,T. (See also Supplementary Material for fitting with other cases such as three-dimensional (3D) metamagnetism and antiferromagnetic fluctuations~\cite{SM}.) The best fit, among all the considered models, is achieved with 2D metamagnetism, suggesting that 2D metamagnetic quantum critical fluctuations are dominanting at high magnetic fields. The lower-bound field of 15 T coincides well with the field of $T_c$($B$) minimum, $B^\star$. Furthermore, above the same field, specific heat~\cite{Rosuel2022} also follows the theoretical power-law, in agreement with our result (see Supplementary Material~\cite{SM}). This indicates that the system at low fields is initially non-critical with the linear-in-field dependence and then crossovers to the two-dimensional quantum critical behavior above 15\,T. Given that the theory addresses $Q=0$ fluctuations~\cite{Zacharias13}, the coincidence between $B^\star$ and the lower field boundary of the quantum critical regime hints at the involvement of $Q = 0$ metamagnetic quantum fluctuations in bolstering the superconducting transition temperature.

While this correlation is suggestive, it should be interpreted with caution until further evidence solidifies this connection, because antiferromagnetic (AFM, $Q\neq 0$) excitations have been identified in neutron scattering experiments as reported in \cite{Duan2021,Knafo2021a,Butch2022}. A key unresolved question concerns how these AFM fluctuations evolve under magnetic fields. The observation that $\Delta S$ without zero-field entropy markedly reflects the anomaly associated with the metamagnetism, underscores the predominant role of metamagnetism at magnetic fields. We note that the change in entropy in our study, represented by $\Delta S$, accounts for merely 10\% of $R\ln(2)$ at its peak. This is substantially smaller compared to the entropy increase with temperature at zero magnetic field, which reaches 80\% of $R\ln(2)$ at 25 K \cite{Willa2021}. Consequently, the anomaly due to metamagnetism become less discernible in the absolute entropy, obscured by the dominant zero-field entropy (See Supplementary Material \cite{SM}).This could be interpreted as at zero (or low) magnetic fields, AFM contributions being more significant, whereas at high magnetic fields, the influence of metamagnetism prevails. Such a trend aligns well with the fundamental behaviour of AFM coupling, which tends to diminish in the presence of magnetic fields. Furthermore, a recent NMR study highlights dominant ferromagnetic fluctuations along the $a$-axis \cite{Fujibayashi2023}. While our findings suggest the dominance of $Q=0$ fluctuations, further direct investigation into magnetic fluctuations with $Q$-resolved experiments, particularly through neutron scattering, is essential for further understanding.
\color{black}

The recent improvements of sample quality have enhanced $T_c$ by 25 \% from the originally reported value~\cite{Sakai2022}. Notably, the upper critical field, $B_{c2}$, is enhanced twice for the field along the magnetic easy $a$-axis, owing to a weak metamagnetism around 8 T~\cite{Tokiwa2023arxiv}. In sharp contrast, $T_c$($B$) curve for the field along $b$-axis does not change its characteristic shape with a minimum at the $B^\star$. Remarkably, $B^\star$ is independent from the sample quality~\cite{Sakai2023}. This is most likely because $B^\star$ is determined by metamagnetic fluctuations, which are not influenced from the crystal quality~\cite{Tokunaga2023}. This assertion is further supported by the fact that the metamagnetic critical field is independent from the sample quality~\cite{Wu2023_arxiv}.

In this study, we show the accelerated increase of entropy with magnetic field along $b$-axis in UTe$_2$ due to $Q=0$ metamagnetic fluctuations. The field dependence in the Fermi liquid regime is well described by Hertz-Millis-Moriya theory for quantum criticality of itinerant metamagnetism. The coincidence between the low-field bound of the quantum-critical fluctuations and $B^\star$ indicates that $Q=0$ fluctuations boost superconductivity. This is in line with the result for the field along the easy $a$-axis, showing that the weak metamagnetism strongly enhances the upper critical field~\cite{Tokiwa2023arxiv}. Thus, our results show that $Q=$0 fluctuations play crucial roles in the spin-triplet superconductivity in UTe$_2$ and promotes further experimental research to more conclusively determine the nature of fluctuations under high magnetic fields. \color{black}

We thank M. Garst, Y. Yanase, H. Kusunose and T. Takimoto for stimulating discussions. We thank M. Nagai and K. Shirasaki for experimental support. The work was supported by JSPS KAKENHI Grant Numbers, JP16KK0106, JP17K05522, JP17K05529, and JP20K03852 and by the JAEA REIMEI Research Program. This work (A part of high magnetic field experiments) was performed at HFLSM under the IMR-GIMRT program (Proposal Numbers 202012-HMKPB-0012, 202112-HMKPB-0010, 202112-RDKGE-0036 and 202012-RDKGE-0084).


\medskip

%

%

\cleardoublepage

\pagebreak
\begin{center}
\textbf{\large Supplemental Materials for "Reinforcement of superconductivity by quantum critical fluctuations of metamagnetism in UTe$_2$"}
\end{center}
\setcounter{equation}{0}
\setcounter{figure}{0}
\setcounter{table}{0}
\setcounter{page}{1}
\makeatletter
\renewcommand{\theequation}{S\arabic{equation}}
\renewcommand{\thefigure}{S\arabic{figure}}
\renewcommand{\bibnumfmt}[1]{[S#1]}
\renewcommand{\citenumfont}[1]{S#1}

\section{Consistency with specific heat}

Figure S\ref{comp}(a) compares specific heat divided by temperature, $C/T$, at 1.86 K \cite{Rosuel2022} and $\int(dM/dT)/TdB$ at 3 K. At low temperatures in Fermi liquid regime, the two quantities are equal, because $\int(dM/dT)/TdB$=$\int(dS/dB)/TdB$=$\gamma$. The two data sets compares well, despite the difference in measurement methods, sample quality and temperature. The  discrepancies can also be attributed to the potential misalignment of the samples to the field direction, which can affect the metamagnetic behaviour. Figure S\ref{comp}(b) demonstrates that when $\int (dM/dT)/TdB$ data are scaled, it closely matches the behaviour of $C/T$, implying a shared functional form between the two. This comparison supports our conclusion that both data sets can be consistently explained by the same model.

\section{Metamagnetic quantum critical behavior in specific heat}

Figure S\ref{CT} shows specific heat divided by temperature of UTe$_2$ for the field along the $b$-axis, taken from Ref.~\cite{Rosuel2022}. The strong enhancement toward metamagnetic field, 34.75\,T, is reminiscent of the other examples of itinerant metamagnet, such as Sr$_3$Ru$_2$O$_7$, YbAgGe, CeRhSn and UCoAl~\cite{Rost-Science09,tokiwa13,Tokiwa-SA15,Tokiwa2016}. This is due to fluctuations caused by metamagnetic quantum critical endpoint (QCEP)~\cite{Zacharias13}. Exactly at QCEP $C$/$T$ diverges. The critical contribution in $C$/$T$ in the diverging regime is $C$/$T\sim b^{(d-3)/3}$, where $b$=$|B$-$B_c|$ is the distance to the critical field, $B_c$, and $d$ is the dimensionality. Since Fermi surfaces consist of mainly cylinders, we assume $d$=2. We fix $B_c$=34.75\,T, as reported~\cite{Rosuel2022}. In Fig.~\ref{CT} we show a fitting result with $C$/$T$=$\gamma$+$ab^{1/3}$, where $\gamma$ and $a$ are fitting parameters. The data follows the theoretical power-law field dependence from 15\,T to 32\,T. The deviation above 32\,T is caused by the fact that the system still has some distance to QCEP. Only a system with a QCEP exhibits a divergence of $C$/$T$ and otherwise it deviates downwards~\cite{Zacharias13}. We note that the data starts to follow the exponent from $\sim$15\,T, in agreement with the entropy increment shown in the main text.

Figure S\ref{CTfit} displays the $C/T$ data as a function of $b$ \cite{Rosuel2022}. This data is analyzed using theoretical models for two-dimensional antiferromagnetic (2D AFM) and two-dimensional metamagnetic (2D MM) quantum criticality. Due to the presence of a constant $\gamma$ term, distinguishing between the models based solely on their slopes is challenging. Nonetheless, the 2D MM model demonstrates a wider range of fitting, which further suggests that metamagnetic fluctuations are likely responsible for the observed anomalous increase in $C/T$.

\begin{figure}
\includegraphics[width=0.8\linewidth,keepaspectratio]{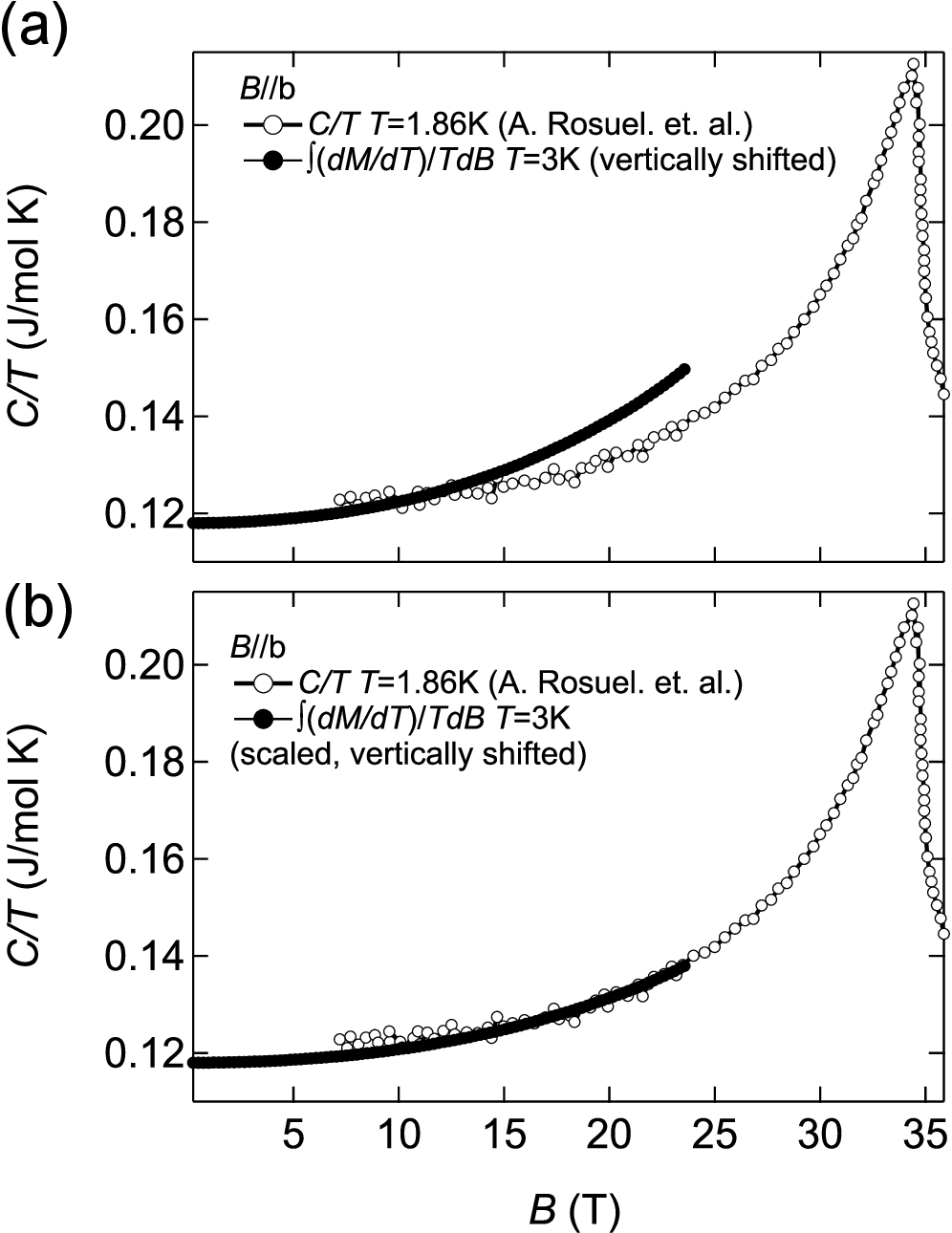}%
\caption{\label{comp} (a) Comparison between $C/T$ at 1.86 \cite{Rosuel2022} K and $\int (dM/dT)/TdB$ at 3 K of UTe$_2$ for magnetic field along the $b$-axis. The latter is shifted vertically. (b) The same data but $\int (dM/dT)/TdB$ is scaled to fit with $C/T$.}
\end{figure}

\begin{figure}
\includegraphics[width=\linewidth,keepaspectratio]{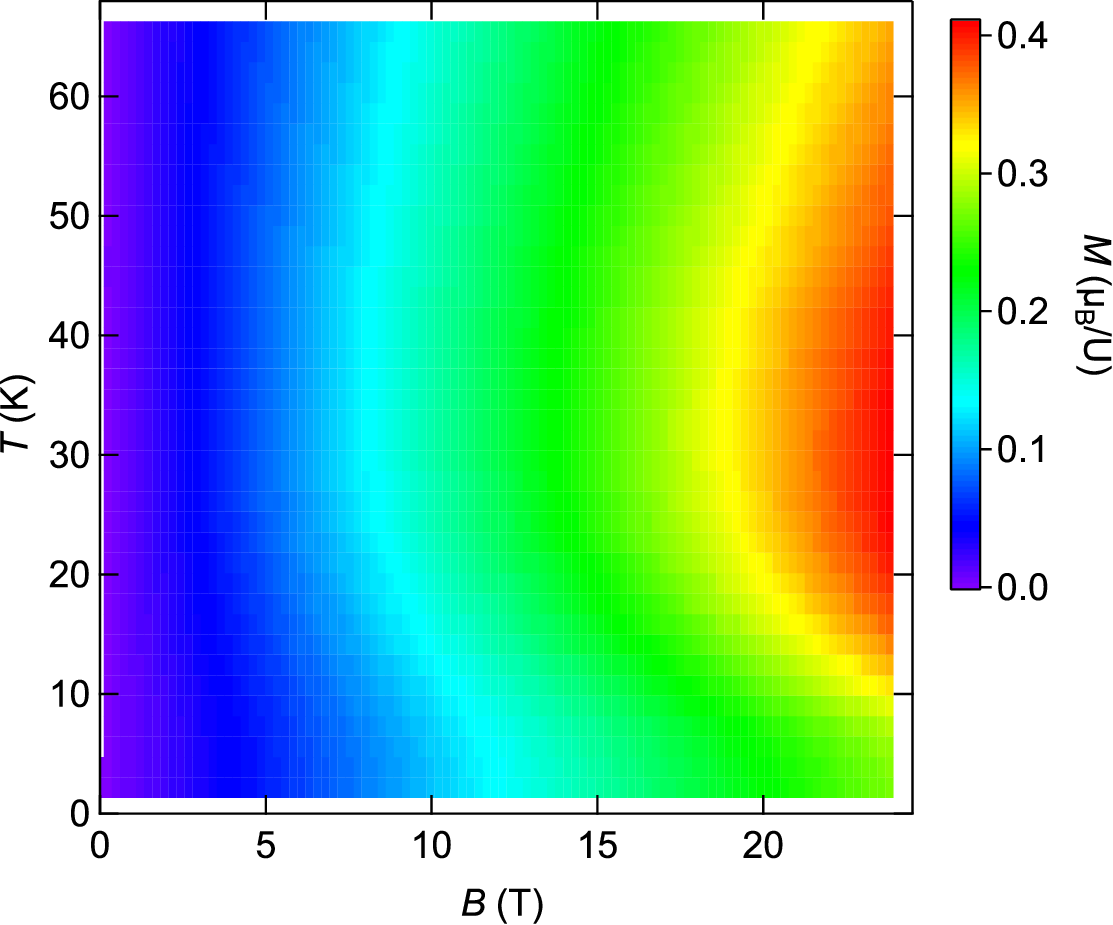}%
\caption{\label{chi_nom}The color contour of magnetization of UTe$_2$ for magnetic field along $b$-axis.}
\end{figure}

\begin{figure}[h]
\includegraphics[width=0.8\linewidth,keepaspectratio]{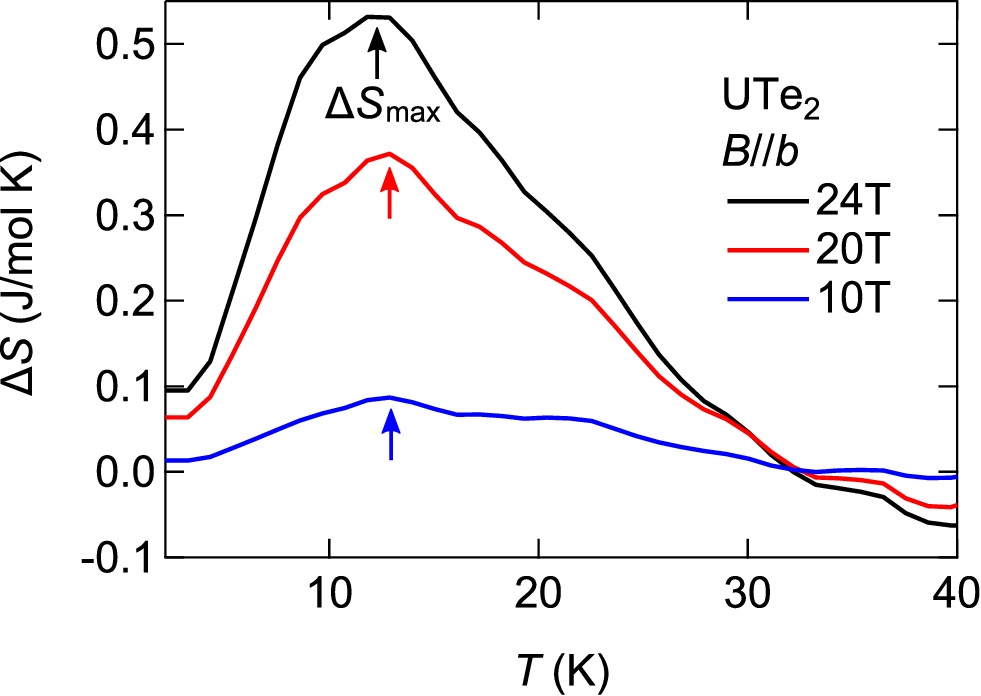}%
\caption{\label{S_Tdep} Entropy increment $\Delta S$ of UTe$_2$ as a function of temperature at different magnetic fields. The arrows indicate a maximum.}
\end{figure}

\begin{figure}
\includegraphics[width=\linewidth,keepaspectratio]{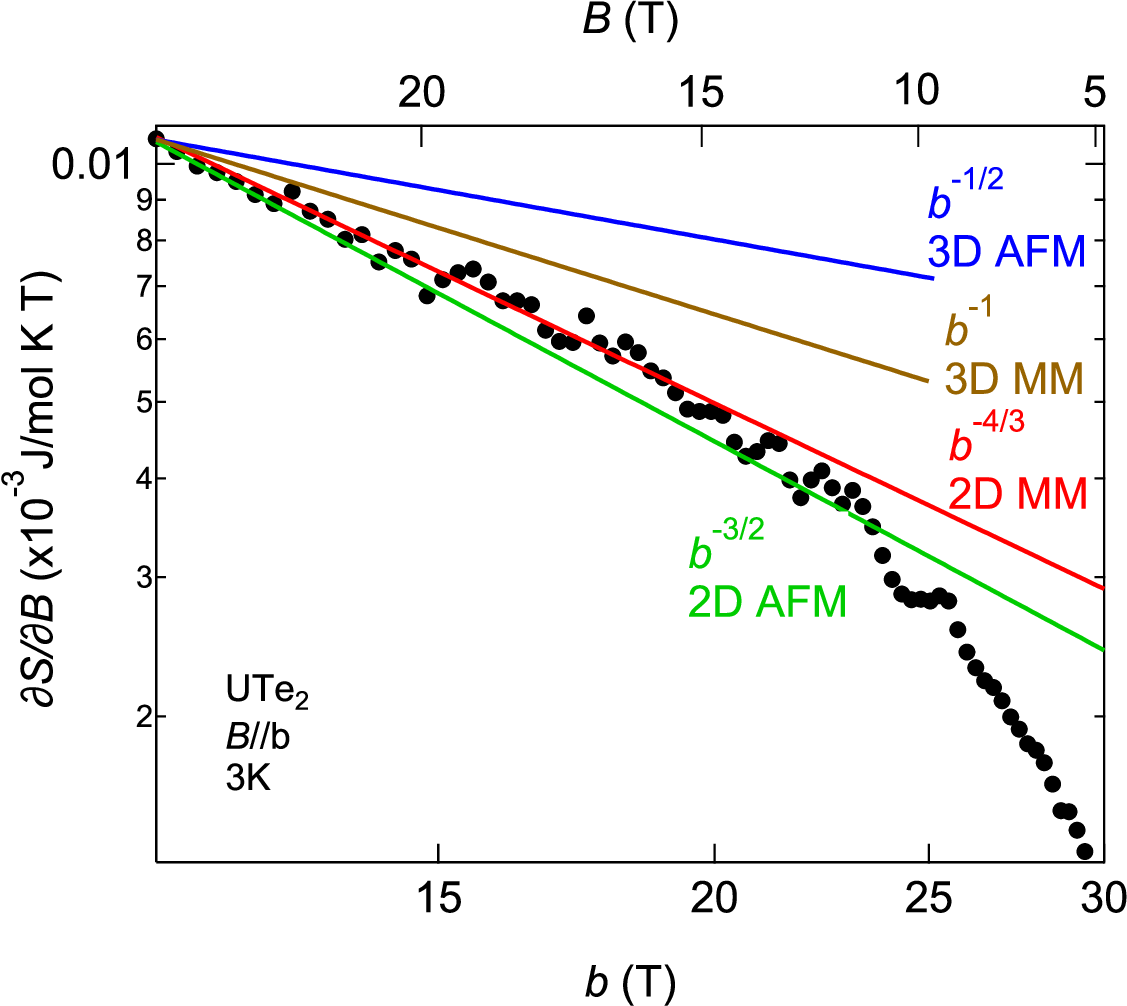}%
\caption{\label{log-log} Log-log plot of the magnetic-field derivative of entropy, $\partial S/\partial BT$, for UTe$_2$ at 3 K, as a function of the distance from the metamagnetic transition field ($B_c = 34.75$ T) along the $b$-axis. The data are the same as the ones shown in Fig. 3(b) in the main text. Various theoretical predictions are depicted by the lines~\cite{millis02, zhu, Zacharias13}. "MM" and "AFM" represent metamagnetism and antiferromagnetism, respectively.}
\end{figure}

\begin{figure}
\includegraphics[width=\linewidth,keepaspectratio]{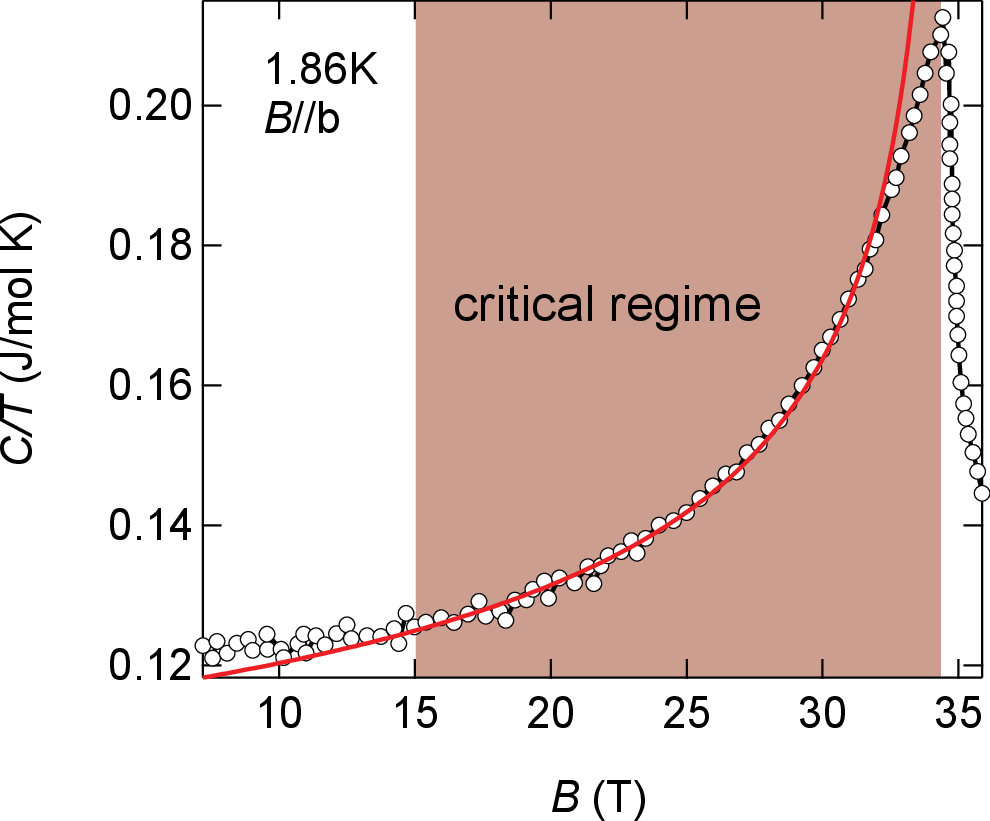}%
\caption{\label{CT}Specific heat data taken from Ref.~\cite{Rosuel2022}. Specific heat divided by temperature, $C$/$T$, is plotted against magnetic field. Red line is a fit to the data by $C$/$T$=$\gamma$+$ab^{1/3}$ \cite{Zacharias13}, where $\gamma$, $a$ are fitting parameters and $b$=$|B$-$B_c|$ is the distance to the critical field, $B_c$=34.75 T \cite{Rosuel2022}. }
\end{figure}

\begin{figure}
\includegraphics[width=\linewidth,keepaspectratio]{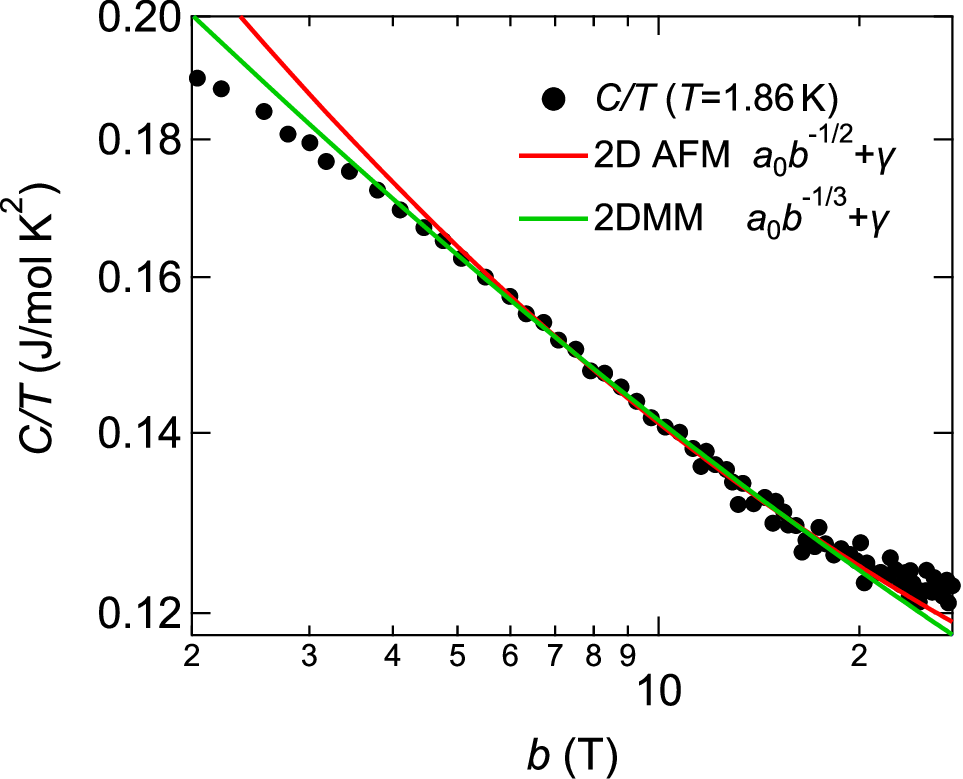}%
\caption{\label{CTfit}Specific heat divided by temperature, $C$/$T$, taken from Ref.~\cite{Rosuel2022} plotted in log scale against the distance to the metamagnetic field, $b=B_c-B$, where $B_c=34.75$ T. The red and green lines are fits to the data by $C$/$T$=$\gamma$+$a_0b^{-1/3}$ and $C$/$T$=$\gamma$+$a_0b^{-1/2}$, where $\gamma$, $a_0$ are fitting parameters \cite{Zacharias13,zhu}. }
\end{figure}

\begin{figure}[H]
\includegraphics[width=\linewidth,keepaspectratio]{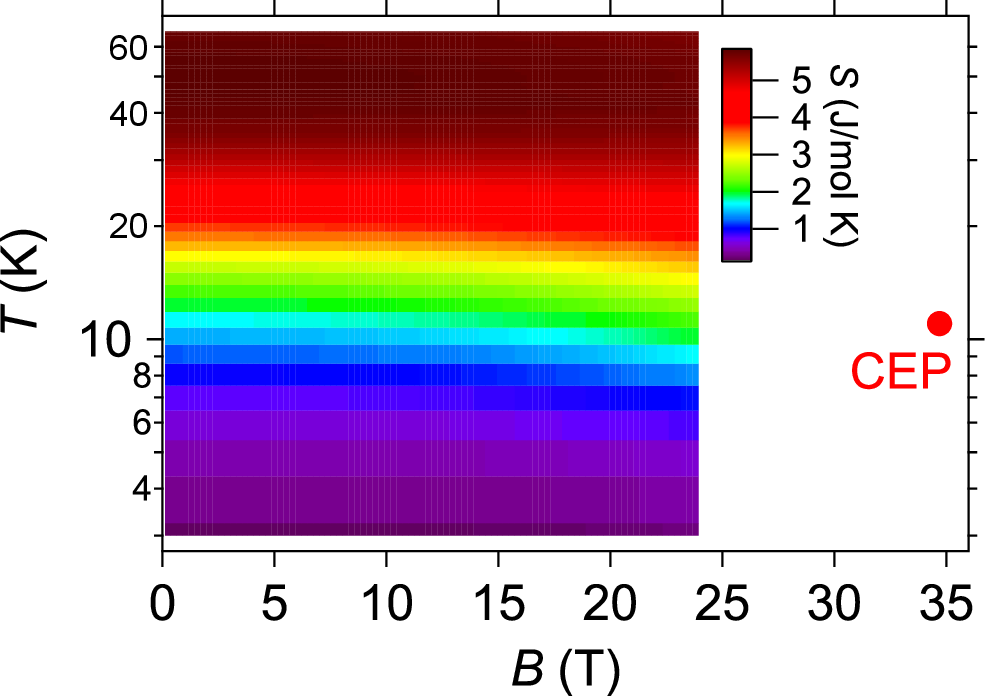}%
\caption{\label{comp} Color contour plot of entropy of UTe$_2$ in the temperature-magnetic field ($\parallel a$-axis) parameter space. The entropy at zero field is from \cite{Willa2021}}
\end{figure}

\end{document}